\documentclass{emulateapj}

\usepackage{times}

\begin{document}
\shorttitle{Morphologies of Old Galaxies at $z\sim1.5$}
\shortauthors{McGrath et al.}
\journalinfo{Accepted for Publication in ApJ}
\submitted{Accepted for Publication in ApJ}

\title{Morphologies and Color Gradients of Luminous Evolved Galaxies at \lowercase{$z\sim1.5$}\altaffilmark{1}}

\author{Elizabeth J. McGrath\altaffilmark{2} and Alan Stockton\altaffilmark{3}}
\affil{Institute for Astronomy, University of Hawaii, 2680 Woodlawn Dr., Honolulu, HI 96822}
\author{Gabriela Canalizo}
\affil{Department of Physics and Astronomy, University of California, Riverside, CA 92521}
\author{Masanori Iye}
\affil{Optical and Infrared Astronomy Division, National Astronomical Observatory of Japan, Mitaka, Tokyo 181-8588, Japan}
\author{Toshinori Maihara}
\affil{Department of Astronomy, Kyoto University, Kitashirakawa-Oiwake-cho, Sakyo-ku, Kyoto 606-8502, Japan}

\altaffiltext{1}{Based on observations made with the 
NASA/ESA Hubble Space Telescope, obtained at the Space Telescope Science 
Institute, which is operated by the Association of Universities for Research in 
Astronomy, Inc., under NASA contract NAS 5-26555.  These observations are 
associated with program \# GO-10418.}
\altaffiltext{2}{Currently at the University of California Observatories/ Lick Observatory, Department of Astronomy and Astrophysics, University of California at Santa Cruz, 1156 High St., Santa Cruz, CA 95064.}
\altaffiltext{3}{Also at Cerro Tololo Interamerican Observatory, Casilla 603, La Serena, 
Chile.}

\begin{abstract}

We have examined in detail the morphologies of seven $z\sim1.5$ passively evolving luminous red galaxies using high resolution HST NICMOS and ACS imaging data.  
Almost all of these galaxies appear to be relaxed systems, with smooth morphologies at both rest-frame UV and visible wavelengths.  Previous results from spectral synthesis modeling favor a single burst of star formation more than 1 Gyr before the observed epoch.  The prevalence of old stellar populations, however, does not correlate exclusively with early-type morphologies as it does in the local universe; the light profiles for some of these galaxies appear to be dominated by massive exponential disks.  This evidence for massive old disks, along with the apparent uniformity of stellar age across the disk, suggests formation by a mechanism better described as a form of monolithic collapse than as a hierarchical merger.  These galaxies could not have undergone a single major merging event since the bulk of their stars were formed, more than 1 Gyr earlier.  There is at least one case, however, that appears to be undergoing a ``dry merger", which may be an example of the process that converts these unusual galaxies into the familiar spheroids that dominate galaxies comprising old stellar populations at the present epoch.  

\end{abstract}


\section{Introduction}
A wealth of observations over the past decade have collectively converged on the $\Lambda$CDM ``concordance" cosmology, 
including results from high-$z$ supernovae \citep{rie98, per99, ton03, woo07}, 
the 2dF Galaxy Redshift Survey \citep{pea01, per02, col05}, 
SDSS \citep{teg04, eis05}, 
WMAP \citep{ben03, spe06}, 
and weak lensing \citep{hoe06, mas07}.
Unfortunately, even given the precision of these measurements and the agreement between such a wide variety of independent observations, the formation of structure on galaxy-size scales remains a challenge for cosmological models.
In particular, the formation and evolution of the first massive galaxies 
 is still poorly understood.  This is distressing given that more than half of all luminous matter in the local universe is contained within massive elliptical galaxies built up at earlier times \citep{fuk98, bel03}.
 
Simple semi-analytic models (SAMs) based on the standard cosmological paradigm predict that massive galaxies form rather late in the history of the universe ($1<z<2$) from successive mergers of smaller galaxies \citep[e.g.,][]{tho99, kau00}.
These models, however, 
have traditionally suffered from an overprediction of the number density of both low and high luminosity galaxies, as well as an inverted color-magnitude relation (i.e., luminous blue galaxies), and an inability to reproduce the observed color bi-modality in galaxy distributions.  
Feedback from sources such as star formation and AGN (e.g., \citealt{gra04, cro06, bow06}) seems necessary in order to reconcile these models with observations.  
Results from the latest SAMs predict that, while the {\it stars} in massive elliptical galaxies formed quickly and early on, their {\it assembly} into the most massive galaxies is characterized by a late, hierarchical build-up \citep{del06}.

The predictions of hierarchical assembly can be explored through high resolution morphological studies of massive galaxies comprising old stellar populations at high redshift.  Examples of such galaxies have become increasingly abundant in the literature \citep{liu00, dad00, dad05, cim02, cim04, iye03, mcc04, yan04, fu05, lab05, sar05, red05, red06, sto04, sto06, sto07, kri06, pap06, abr07, mcg07}; however morphological 
information has still been relatively sparse.  It is commonly assumed that spheroids 
should dominate evolved galaxies at high redshift as they do in the local universe.  However, some recent studies have found evidence for S0-type galaxies as well as exponential disks among these evolved galaxies at high $z$ \citep{iye03, cim04, yan04, dad05, fu05, sto04, sto06, sto07}.
We present a sample of seven $z\sim1.5$ ``old galaxies'' (OGs, by which we refer to the age of the dominant stellar population and not necessarily the age given by the time 
elapsed since galaxy assembly) selected to have both $R-K^{\prime} > 6$, and $J-K^{\prime}\sim2$.   
We have obtained high-resolution morphological data for these objects from HST ACS and NICMOS imaging.  
We discuss the range in morphological classes among OGs in our sample at $z\sim1.5$ and the implications for galaxy formation scenarios.  
Throughout this paper we assume a flat cosmology with $H_0 = 70$ km s$^{-1}$ Mpc$^{-1}$, $\Omega_M =0.3$ and $\Omega_{\Lambda}=0.7$.

\section{Observations}
In \citet*[hereafter Paper I]{mcg07}, 
we presented six old galaxy candidates in the fields of radio-loud QSOs at $z=1.5$, five of which were dominated by stellar populations that were already $\gtrsim 1$ Gyr old.  Here we continue our study of these six objects, with high resolution imaging from HST.  In addition, we include a seventh source, J094258+4259.2 ERO, discovered serendipitously in the field of Abell cluster 851 by \citet{iye00, iye03}.

Each galaxy was observed with HST at approximately rest-frame U and R-bands using the ACS F814W filter (2 orbits) and the NICMOS2 F160W filter (1 orbit).  Details of the
observations are given in Table 1 of Paper 1, except for J094258+4659 ERO, which
was observed on 2005 April 18 (UT) with ACS for 5464 s and
on 2005 April 19 (UT) with NICMOS2 for 2880 s.
We used the standard pipeline-processed ACS images for analysis.  The NICMOS images, however, required a more careful processing.  We first checked to make sure the observations were taken long enough after SAA passage ($>30$ min.) so that they did not suffer from cosmic ray persistence.  We then took the \emph{calnica} processed images, which had been flat-fielded and corrected for amplifier glow, as a starting point and corrected for bias offsets and inverse flatfield effects using STSDAS tool \emph{pedsky}.  We generated a bad pixel mask from the data quality file, and included an additional mask for the coronagraphic occulter which produces background radiation detected in the F160W filter.  We used \emph{L.~A.~Cosmic} \citep{van01}
to perform a first cut of cosmic rays from the individual images, and then drizzled the individual dither positions together to obtain a cosmic-ray-free, subsampled image.  We performed a series of tests with Tiny~Tim\footnote{http://www.stsci.edu/software/tinytim/tinytim.html}
generated PSFs at the location of our galaxies on the NIC2 chip in each of the dither positions to determine the best drizzle parameters required to maintain a tight PSF as well as recover some resolution in the final image.  The parameters we adopted were an input drop size of 0.7 convolved with a gaussian kernel and an output pixel size of 0.5 times the original.

\section{Stellar Population Properties}

\tabletypesize{\scriptsize}
\begin{deluxetable}{rccccc}
\tablecolumns{6}
\tablewidth{0pc}
\tablecaption{Summary of Simple Stellar Populations}
\tablehead{
\colhead{Galaxy} & \colhead{$z_{spec}$} & \colhead{Age} & \colhead{A$_V$} & \colhead{P($\chi^2$)} & \colhead{Stellar Mass} \\
\colhead{} & \colhead{} & \colhead{(Gyr)} & \colhead{(mag)} & \colhead{\%} &  \colhead{(10$^{11} M_{\sun}$)}
}
\startdata
TXS\,0145+386 ER1 & 1.4533 & 0.90 & 0.79 & 70 & 1.74 \\
ER2 & 1.459\tablenotemark{$\ast$} & 2.40 & 0.02 & 61 & 2.02 \\
J094258+4659 ERO\tablenotemark{$\dagger$} & 1.5\tablenotemark{$\ast\ast$} & 1.0 & $\lesssim 1.4$ & \nodata & 3.0 \\
B2\,1018+34 ER2 & 1.4057 & 0.64 & 0.94 & 6 & 1.42 \\
TXS\,1211+334 ER1 & 1.598 & 1.43 & 0.23 & 91 & 2.90 \\
4C\,15.55 ER2 & 1.412 & 1.80 & 0.02 & 89 & 1.75 \\
TXS\,1812+412 ER2 & 1.290 & 1.43 & 0.51 & 44 & 1.56 \\
\enddata
\label{tab-sp}
\tablenotetext{$\ast$}{The best-fit photometric redshift is given as no spectroscopy was obtained for this source.}
\tablenotetext{$\dagger$}{Data are from \citet{iye03} and were determined using BC03 rather than CB07 models.}
\tablenotetext{$\ast\ast$}{The weighted best-fit photometric redshift is given since no prominent absorption or emission features were visible in the rest-frame visible spectrum of \citet{iye03}.}
\end{deluxetable}

Stellar populations for six of these galaxies were presented in Paper I and were determined from broadband photometric fits to S. Charlot \& G. Bruzual (2007, private communication; hereafter CB07) instantaneous burst models.  In that paper we discussed uncertainties in the models and how the age-metallicity degeneracy affects our resulting age determinations.  Extremely deep imaging was obtained at blue wavelengths in order to place constraints on the rest-frame UV portion of the spectral energy distribution (SED).  This spectral region is important for age-dating these galaxies, since even a small contribution by mass from young stellar populations would cause a disproportionate increase in the total light at these wavelengths.  For this reason, rest-frame UV spectra for five of these six sources were also obtained and were fit to a variety of models, including the CB07, \citet{bru03}, and \citet{mar05} models.  The results from independent fitting of the spectra and the photometry were consistent to within $\sim$100 Myr in most cases (see Paper I).  Furthermore, these spectra provide confirmation of the galaxy redshifts, removing a free parameter from the photometric SED fitting and resulting in more accurate age determinations.  The stellar population properties of the remaining source, ERO J094258+4659.2, were discussed in detail in \citet{iye03}, who used both broadband photometry and a low $S/N$ rest-frame visible spectrum to place limits on the age of the dominant stellar population.  In all but one galaxy in our sample (B2 1018+34 ER2), the dominant stellar populations were found to be $\sim$1 Gyr or older.  A summary of the properties of the stellar populations 
of these sources is given in Table~\ref{tab-sp}.

The use of instantaneous burst models 
places a lower limit on the age of the dominant stellar population.  
More realistic models (e.g., exponentially decaying star formation rates) would require star formation to begin at even earlier times in order to reproduce the observed spectral energy distributions (SEDs), especially the sharp inflection at 4000 {\AA} that is the signature of old stellar populations.  
However, 
given that an instantaneous burst of $\sim10^{11}$M$_{\sun}$ in stars is not strictly physical, 
it is useful to determine the maximum allowable contribution to the SED from more recent episodes of star formation.
For TXS\,0145+386 ER1 and B2\,1018+34 ER2, where we see [\ion{O}{2}] emission in the rest-frame near-UV spectra, if we assume this emission is due purely to star formation, rather than AGN or LINER activity, the rate of current star formation is approximately $2 M{_\sun}$ and $3.5 M{_\sun}$ yr$^{-1}$, respectively (see \citealt{sto06,mcg07}).
For galaxies without [\ion{O}{2}] emission, we can 
place limits on the amount of current star formation directly from the SED modeling.  Assuming a composite stellar population consisting of an instantaneous burst in which the majority of the stars were formed plus a small contribution from an underlying continuous star forming population, we can determine the maximum amount of current star formation that is plausible given the broad-band colors in the SED.  We have re-fit each of our galaxies from Paper I with this model, allowing both the age of the instantaneous burst and the percent contribution from continuous star formation to vary, and we show the results in Table \ref{tab-comp}.  In summary, ages for all the galaxies increase (compared with the pure instantaneous burst model), and star formation rates ${\gtrsim}1 M_{\sun}$yr$^{-1}$ are excluded for every galaxy except B2\,1018+34 ER2.

\begin{deluxetable}{rccccc}
\tablecolumns{6}
\tablewidth{0pc}
\tablecaption{Composite Stellar Population Properties}
\tablehead{
\colhead{Galaxy} & \colhead{Age} & \colhead{A$_V$} & \colhead{current SFR} & \colhead{P($\chi^2$)} & \colhead{Stellar Mass} \\
\colhead{} & \colhead{(Gyr)} & \colhead{(mag)} & \colhead{($M_{\sun}$ yr$^{-1}$)} & \colhead{\%} & \colhead{(10$^{11} M_{\sun}$)}
}
\startdata
TXS\,0145+386 ER1 & 1.02 & 0.75 & 0.85 & 47 & 1.70 \\
ER2 & 2.75 & 0.00 & 0.1 & 12 & 2.15 \\
B2\,1018+34 ER2 & 1.90 &0.29 & 1.0 & 75 & 1.88 \\
TXS\,1211+334 ER1 & 2.10 & 0.00 & 0.2 & 52 & 3.31 \\
4C\,15.55 ER2 & 2.00 & 0.00 & 0.1 & 67 & 1.86 \\
TXS\,1812+412 ER2 & 1.68 & 0.46 & 0.1 & 47 & 1.73 \\
\enddata
\label{tab-comp}
\end{deluxetable}

\section{Morphologies of Old Galaxies at $z\sim1.5$}

\begin{figure*}[b]
\epsscale{1.0}
\plotone{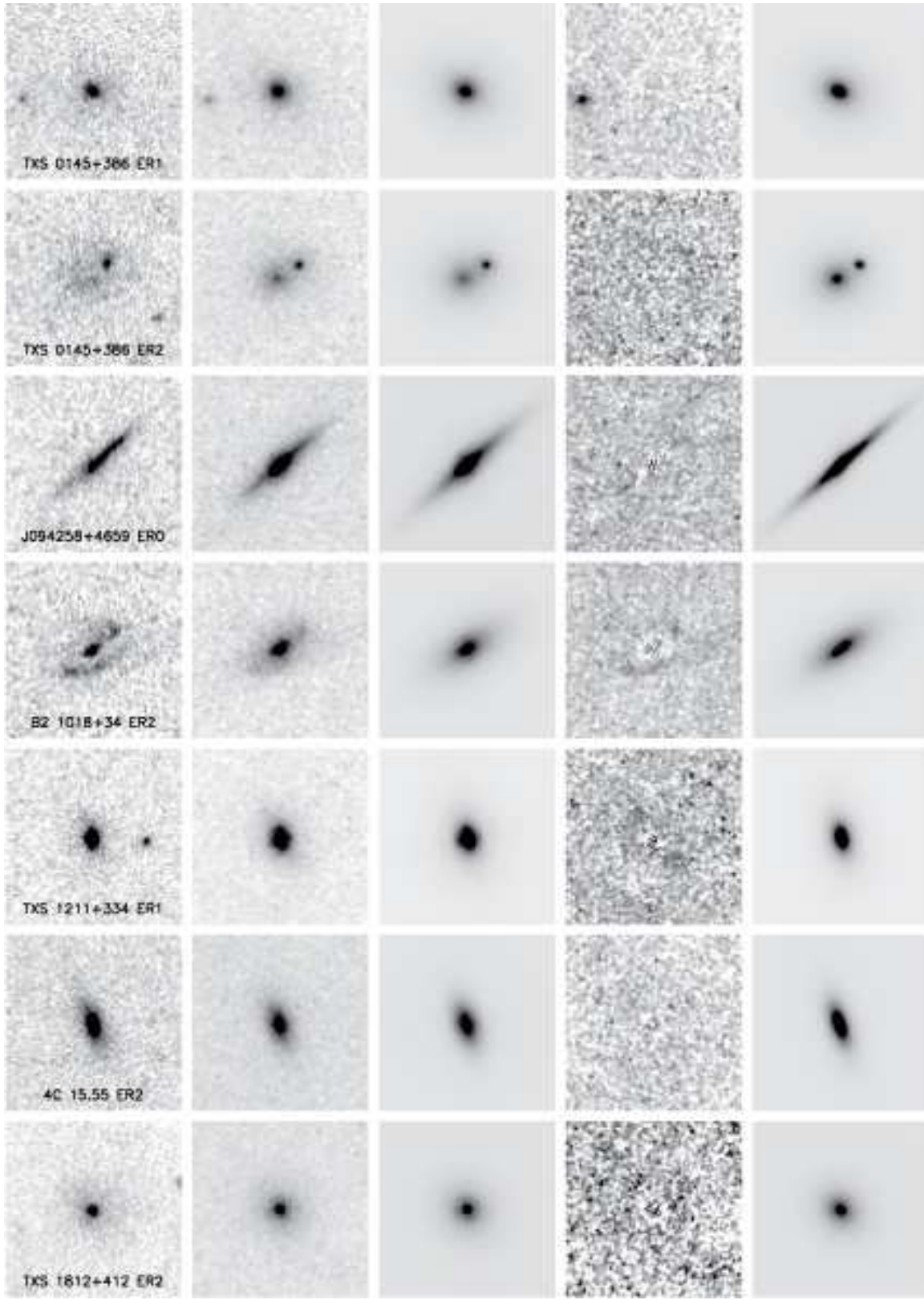}
\caption{HST imaging of old galaxies.  Images are 4$^{\prime\prime}$ on a side, North is up, and East is to the left.  From left to right we show the ACS F814W, NIC2 F160W, best fit {\sc galfit} model for the NICMOS image, residuals obtained from subtraction of the model from the NICMOS image, and the {\sc galfit} model without PSF convolution.  \label{hst_fig}}
\end{figure*}

In order to study the morphologies of these seven galaxies, we used our NIC2 imaging data, 
which have a higher S/N than the ACS images and are less likely to be affected by any small scale or local variations in the light profile.  These images should provide the smoothest galaxy profiles which will in turn provide the best estimate of the underlying galaxy structure.  We then compared the NICMOS profiles with those obtained from the lower S/N ACS images in order to determine if there is any evidence for 
a difference in the distribution of 
older and younger stellar populations 
(whose contributions dominate the light profile long-ward and short-ward of the 4000{\AA}-break, respectively).

We used the program, {\sc galfit}\footnote{\protect{\sc galfit} is available at http://www-int.stsci.edu/$\sim$cyp/work/galfit/galfit.html} \citep{pen02} to model the galaxy light profiles.  These galaxy models were convolved with the Tiny~Tim NICMOS PSF, which was drizzled in the same manner as the observations.  For all galaxies, we determined the best fit exponential, r$^{1/4}$-law, and S\'ersic \citep
{ser68} profiles.  If these simple models did not provide an adequate fit to the data, we
then tried models with a central bulge plus extended disk, or models with a small point-like core.  

PSFs were also generated for the ACS imaging in order to convolve them with a model galaxy to fit the ACS light profiles.  These PSFs required a little more care in making because of the distortion in the ACS field.  Tiny~Tim produces PSFs for ACS that are geometrically distorted, such that they approximate the PSF at the location of an object in the raw data before pipeline processing.  In order to use these PSFs with the final drizzled and distortion-corrected images, one needs to correct these PSFs in the same manner as the data.  One cannot simply use the pre-distorted PSFs from step two of Tiny~Tim because the effects of charge diffusion are not included until the final step and the process of correcting for the distortion could produce artifacts or other imperfections not represented in this idealized PSF.  We used the IDL wrapper for Tiny~Tim created by \citet{rho07}\footnote{see http://www.astro.caltech.edu/$\sim$rjm/acs/PSF/} in order to produce distortion corrected PSFs at the position of our targets, which we then drizzled in the same manner as the data to produce the most realistic representation of the PSF for our ACS imaging.

\begin{figure*}[htb]
\epsscale{0.8}
\plotone{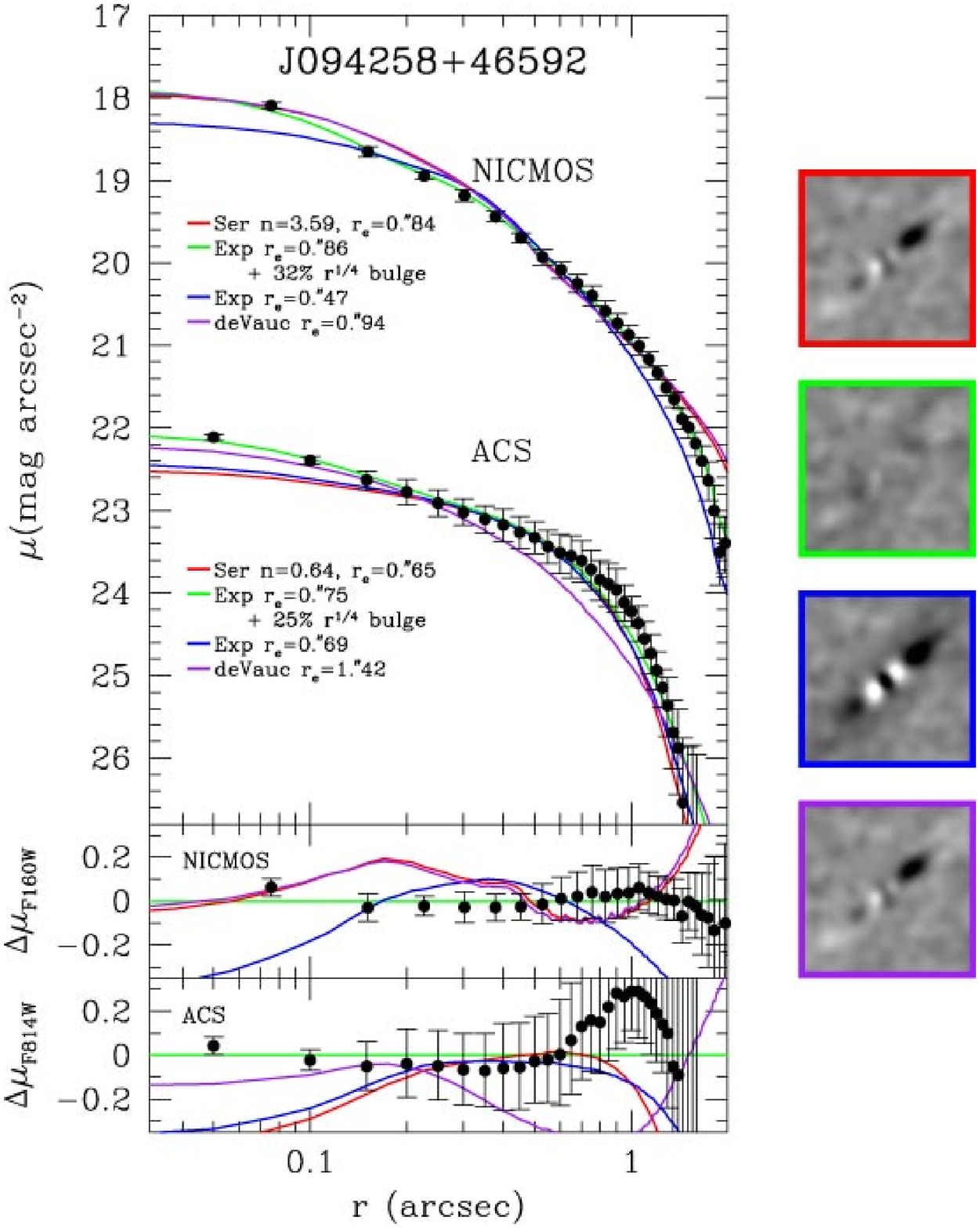}
\caption{Surface brightness profile for J094258+4659 ERO.  
The best fit model is an exponential disk with an r$^{1/4}$ bulge that contributes 32\% of the total light at rest-frame R-band and 25\% of the light at rest-frame U-band. \label{j094258morph}}
\end{figure*}

The observed HST ACS and NICMOS images are shown in Figure \ref{hst_fig} along with the best-fit {\sc galfit} models, the residuals from subtracting the model from the NICMOS image, and the models without PSF convolution, which are representations of the true galaxy light profiles.  The observed surface brightness profiles (Figs. \ref{t0145er1morph} $-$ \ref{t1812morph}) are plotted at radii increasing by intervals of one original (non-subsampled) pixel (0\farcs075 for NICMOS and 0\farcs05 for ACS) outward from the center of the galaxy.  They are averaged along ellipses of constant position angle but varying ellipticity.  Model profiles were determined using the same set of ellipses as those used for the observed data.  Adjacent data points and their errors are correlated to some extent because of PSF smearing and the fact that the ellipses sample some of the same pixels along the minor axis.  A summary of the best-fit morphological properties for each galaxy is listed in Table \ref{tab_prop}.

\subsection{TXS 0145+386 ER1 and ER2}

\begin{figure*}[htb]
\epsscale{0.8}
\plotone{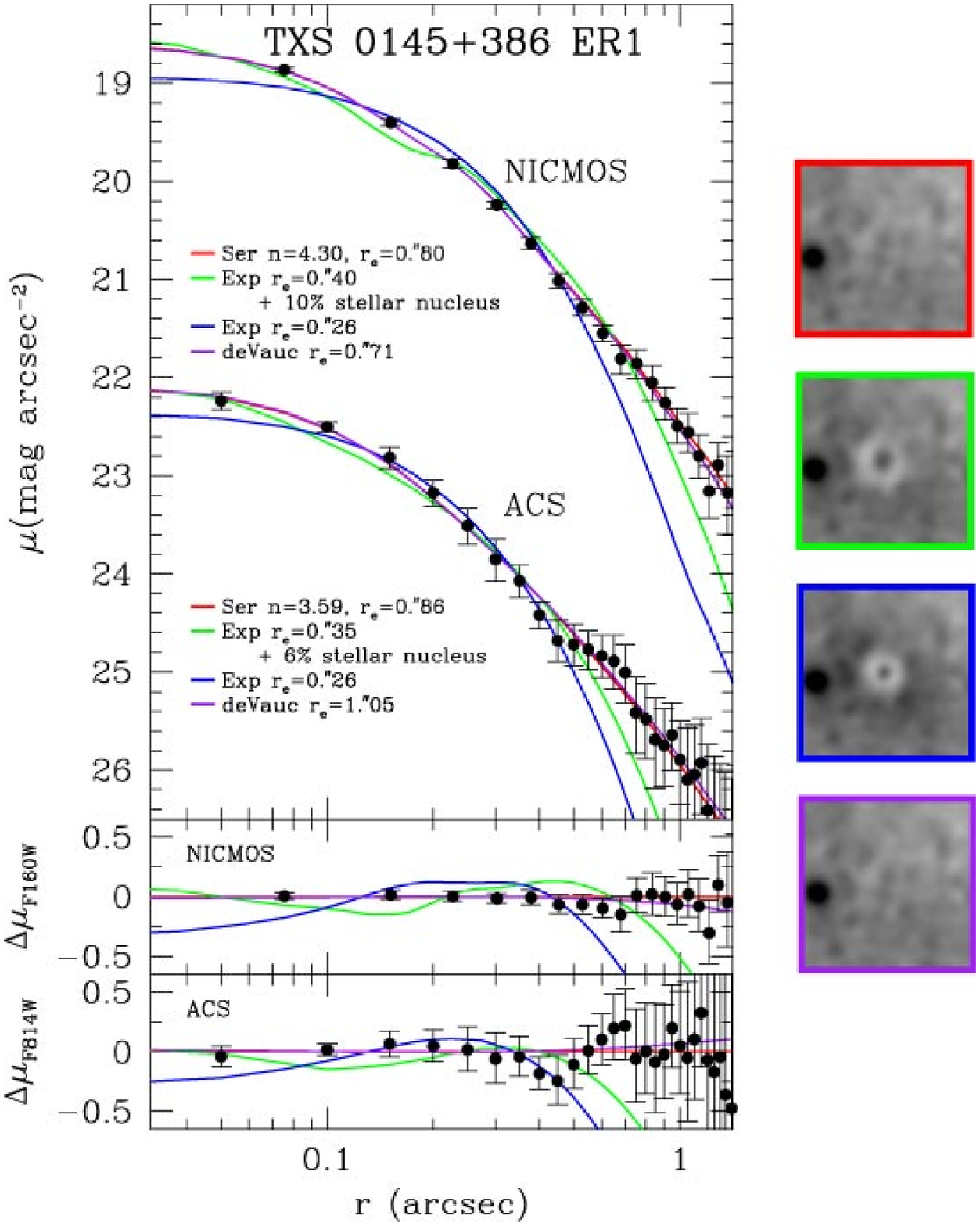}
\caption{Surface brightness profile for TXS 0145+386 ER1.  Data points in this and the following six plots  (Figures 2-8) are from the NICMOS F160W and the ACS F814W observations, and are shown with 3$\sigma$ errors.  Key for the models is given in the figure.  The bottom two panels show the deviation of the data points and other models from the best fit model for both the NICMOS (top) and ACS (bottom) observations.  Residuals after model subtraction from the NICMOS image are shown to the right of the plots and are smoothed by a Gaussian with width, $\sigma$ = FWHM of the PSF to eliminate pixel-to-pixel variations.  These residual images are outlined in color according to the appropriate model, as given in the figure key.  
For TXS\,0145+386 ER1, the S\'ersic and de~Vaucouleurs models overlap for all but the largest of radii, and provide equally good fits to the data.  
\label{t0145er1morph}}
\end{figure*}

\begin{figure*}[htb]
\begin{center}
\epsscale{0.8}
\plotone{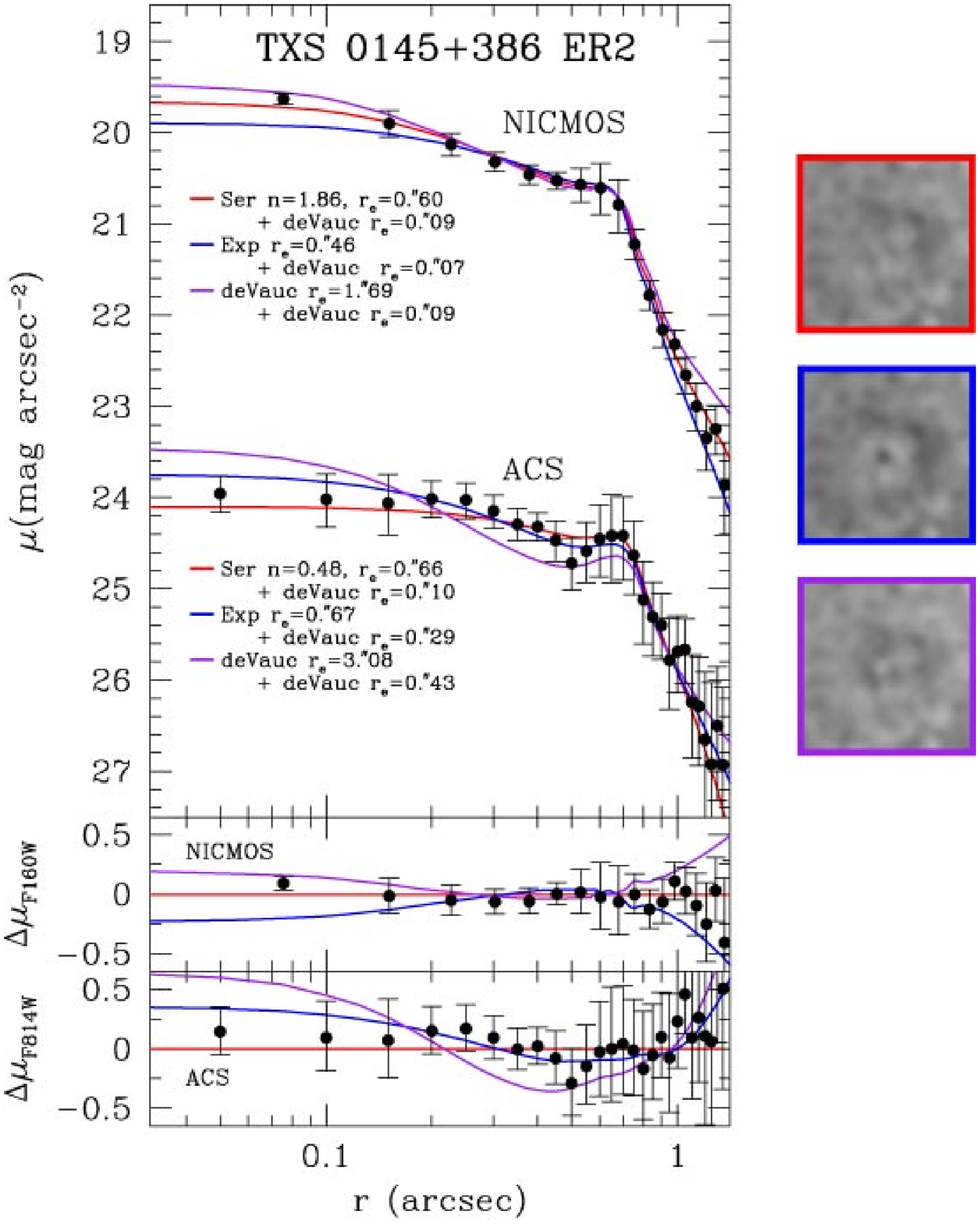}
\end{center}
\caption{Surface brightness profile for TXS 0145+386 ER2.  The profile is centered on the more extended galaxy in this merging pair, and the bump at $\sim$0\farcs7 is due to the second, more compact galaxy.  Errors are determined by the standard deviation from the azimuthal average, and are therefore larger near the 0\farcs7 bump since this is an asymmetric contribution to the flux.  For this source, the two model components listed are simply for each galaxy in the pair, rather than a multicomponent fit to a single galaxy.  The fit to the compact galaxy was held constant with a de~Vaucouleurs profile, while models were allowed to vary for the second, extended galaxy.  Something close to an exponential or low S\'ersic index gives the best fit for this extended galaxy.  A de~Vaucouleurs profile appears to be ruled out by the ACS data.
\label{t0145er2morph}}
\end{figure*}

Two EROs with old stellar populations were found in the field of TXS 0145+386.  From the HST imaging, ER1 appears to be a fairly regular elliptical galaxy.  Previous results from ground-based adaptive-optics (AO) imaging indicated that this galaxy may have been a disk 
galaxy with a low S\'ersic index \citep{sto06}.
However, the S/N of the AO imaging was not very high, particularly in the outermost regions of the galaxy where contribution from the near-IR background becomes important and is both higher and less stable from the ground than it is from space.  It is at large radii where the best-fit exponential and r$^{1/4}$-law profiles deviate most noticeably from each other.
 From the higher S/N NICMOS data, it is clear that a S\'ersic profile with an index $n=4.29$, or very close to a classic de~Vaucouleurs elliptical \citep{deV48}, is the best fit to the observed data (Fig.~\ref{t0145er1morph}), rather than an exponential disk.  The residuals from the fit are within the background noise.  

\citeauthor{sto06}\ noted from their AO imaging that there appeared to be some low surface brightness emission in the outermost regions of the galaxy, ER1.
This is resolved in our ACS and NICMOS images and appears to be a faint, compact, neighboring galaxy.  
There does not appear to be any visible interaction between this faint galaxy approximately 2\arcsec\ (17 kpc) to the east of ER1 and ER1 itself, although the two galaxies have nearly identical colors.  At this redshift, however, particularly given the low luminosity of the ``companion" galaxy, it would be difficult to see the signatures of interaction if the two were undergoing a ``dry merger" since these features would have a low surface brightness and would be dominated by old stars, rather than bright knots of star formation.  We therefore cannot say whether a merger between these two sources is occurring, however, we note that the $H_{F160W}$-band flux ratios of the galaxies implies that any merger would be minor and would not significantly disrupt the structure of the primary galaxy.

ER2, the second old galaxy found in the field of TXS\,0145+386, was unresolved from ground-based, non-AO imaging.  From the HST imaging it is clear that ER2 is in fact an interacting galaxy pair.  One of the galaxies in the pair appears to be much redder  and more extended than the other, more compact galaxy.  Modeling of the galaxy pair in ER2 is a bit more uncertain than modeling a single galaxy due to the increased degrees of freedom.  Nevertheless, the compact galaxy in this pair is well modeled by a S\'ersic profile with a high index ($n\ge4$) and a small effective radius.  We constrained our fit to the $n=4$ case, and only allowed {\sc galfit} to model the second, more extended galaxy.  This second galaxy is well represented by a profile that has a S\'ersic index intermediate between a classical elliptical and a classical disk, although closer to a disk.  The best-fit S\'ersic index gives $n=1.86$ (Fig.~\ref{t0145er2morph}).  The ACS image for this galaxy shows a highly irregular structure, and can only be modeled with a relatively broad and flat profile, similar to that of a low surface brightness disk.

\subsection{J094258+4659 ERO}
J094258+4659 ERO is visibly elongated in both the ACS and NICMOS images, and 
appears to be either an edge-on disk or S0 galaxy with a small central bulge.  Comparison of the 
NICMOS and ACS images shows that the stars in the bulge appear to be somewhat redder than stars in the disk, as expected for typical disk or S0 galaxies.  It is likely that the bulge
component formed first, with the disk completing its star formation only slightly later.  
The dominant stellar population, however, appears to be $>1$ Gyr old (Iye et al.~2003), so star formation in both the disk and bulge components must have occurred extremely rapidly in order to produce such an evolved population by $z=1.5$.

\citet{iye03} noted from their Subaru imaging that J094258+4659 ERO appeared to be an S0-type galaxy with approximately 30\% of the total light being contributed from a central bulge component.  The NICMOS imaging confirms this result and the best fit {\sc galfit} model is an exponential disk with a 32\% bulge component (Fig.~\ref{j094258morph}).  The ACS image, while noisier in the outer regions, also appears to favor a disk-dominated galaxy with 25\% of the total light contributed by a central bulge.

\subsection{B2 1018+34 ER2}

\begin{figure*}[htb]
\begin{center}
\epsscale{0.8}
\plotone{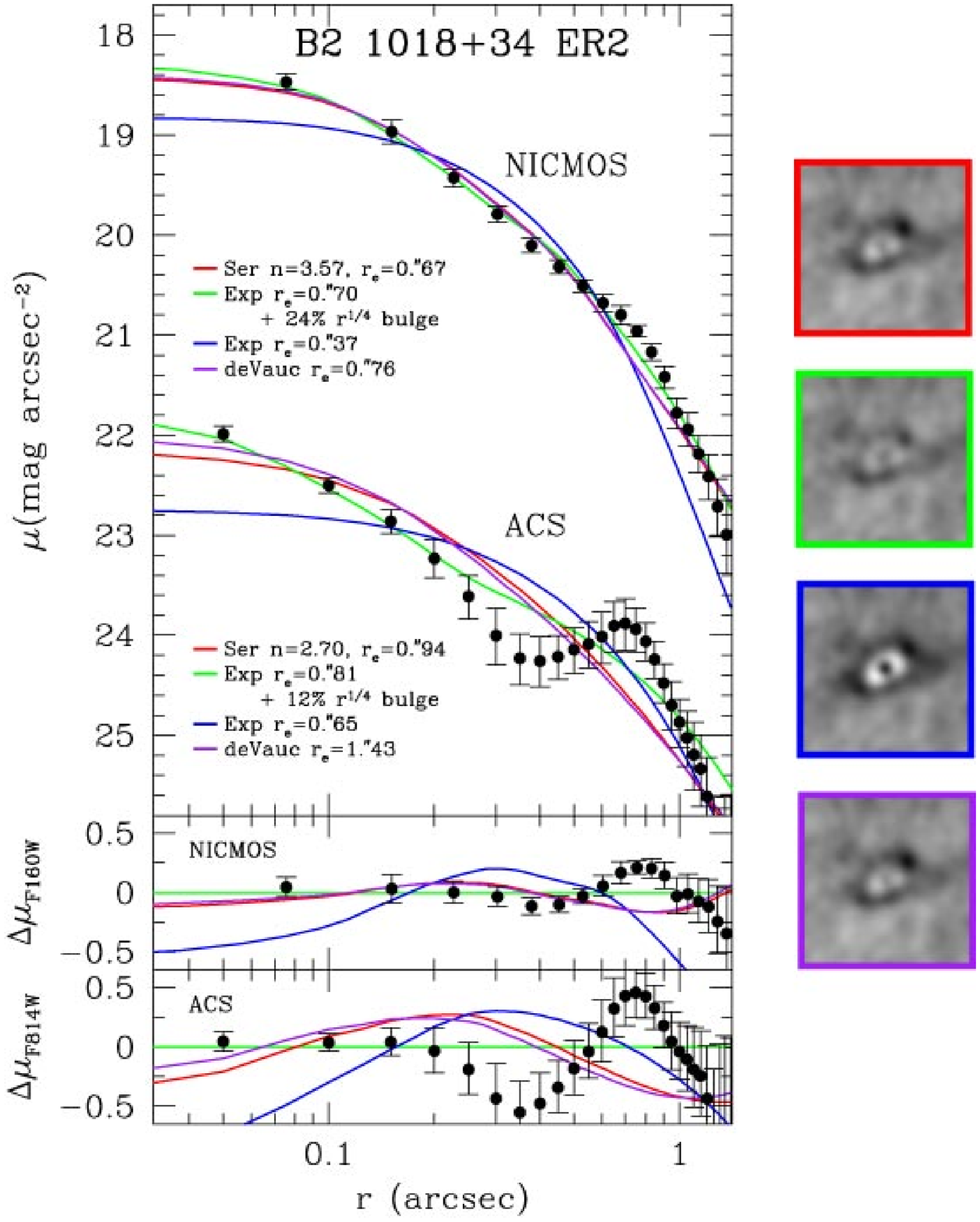}
\end{center}
\caption{Surface brightness profile for B2 1018+34 ER2.  
The best fit model is an exponential disk with an r$^{1/4}$ bulge that contributes 24\% of the total light at rest-frame R-band and 12\% of the light at rest-frame U-band. Spiral arms contribute to radial asymmetries that are difficult to model and are most readily apparent in the ACS profile.  \label{b1018morph}}
\end{figure*}

Inspection of the HST images for B2 1018+34 ER2 reveals a striking grand design spiral arm structure.  The arms are clearly bluer than the central region, especially a few noticeable knots of star formation in the ACS image.  Some of these knots are also visible in the NICMOS image, notably the knots that appear to be on the ends of a possible bar, or the innermost region of the spiral arms.  In 
Paper I 
we concluded that this galaxy consisted of a younger stellar population than the rest of the galaxies in our sample, and noted the presence of [\ion{O}{2}] emission in the rest-frame UV spectrum.
Asymmetries such as spiral arms are not accounted for in the {\sc galfit} modeling, however, a smooth exponential profile with the addition of a central bulge that contributes 24\% of the total light fits the underlying structure of the galaxy well (Fig.~\ref{b1018morph}).  After subtraction of the model galaxy, arm structure that was not as evident in the NICMOS image as it was in the ACS image becomes clearly visible in the residuals.

\subsection{TXS 1211+334 ER1}

\begin{figure*}[htb]
\epsscale{0.8}
\plotone{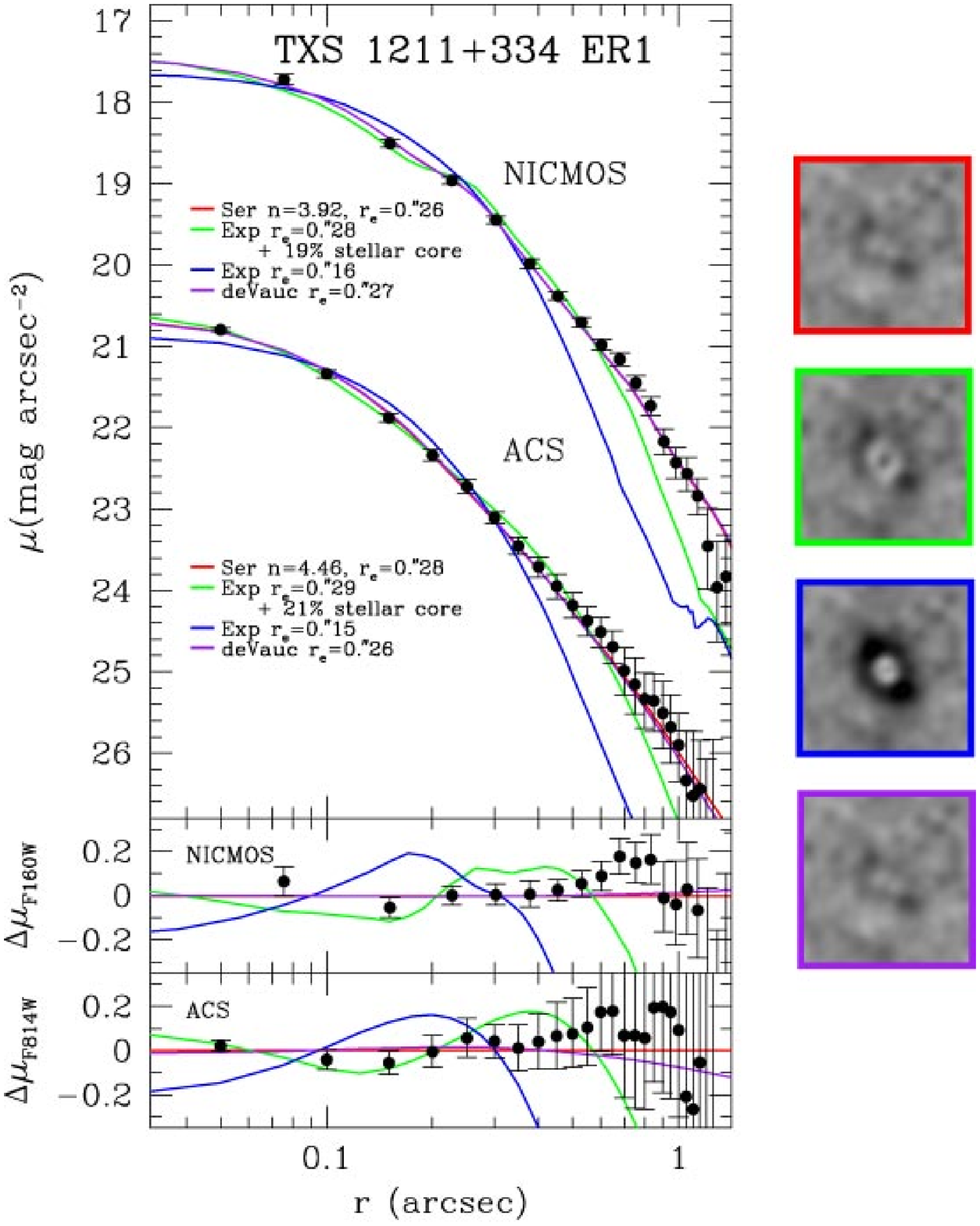}
\caption{Surface brightness profile for TXS 1211+334 ER1.  
The best fit model is a S\'ersic profile with an index $n=3.90$, or a classic de~Vaucouleurs profile. The two models overlap at nearly all radii.  \label{t1211morph}}
\end{figure*}

TXS 1211+334 ER1 appears to be a somewhat elongated elliptical galaxy.
The best fit {\sc galfit} profile is a S\'ersic $n=3.90$ with a small effective radius (Fig.~\ref{t1211morph}).  An exponential profile with a small effective radius plus a central PSF also fits the central regions of the galaxy, but is too low at large radii.  It is possible that this galaxy is still in the process of relaxation into a classical elliptical galaxy.  The residuals from the S\'ersic fit show there is excess flux along the axis of elongation (see Fig.\ref{hst_fig} and residual images in Fig.~\ref{t1211morph}), possibly remnants of a much earlier merging event.  If there was a major merging event in the past, it could not have produced much star formation since the age of the dominant stellar population, as shown by both broad-band photometry and medium-resolution spectra 
(Paper I), 
is extremely old ($>1.4$ Gyr), with little evidence for any admixture from a younger, UV-dominated population.  
The ACS profile is in agreement with the NICMOS profile, also favoring a classical de~Vaucouleurs elliptical galaxy.

\subsection{4C 15.55 ER2}

\begin{figure*}[htb]
\epsscale{0.8}
\plotone{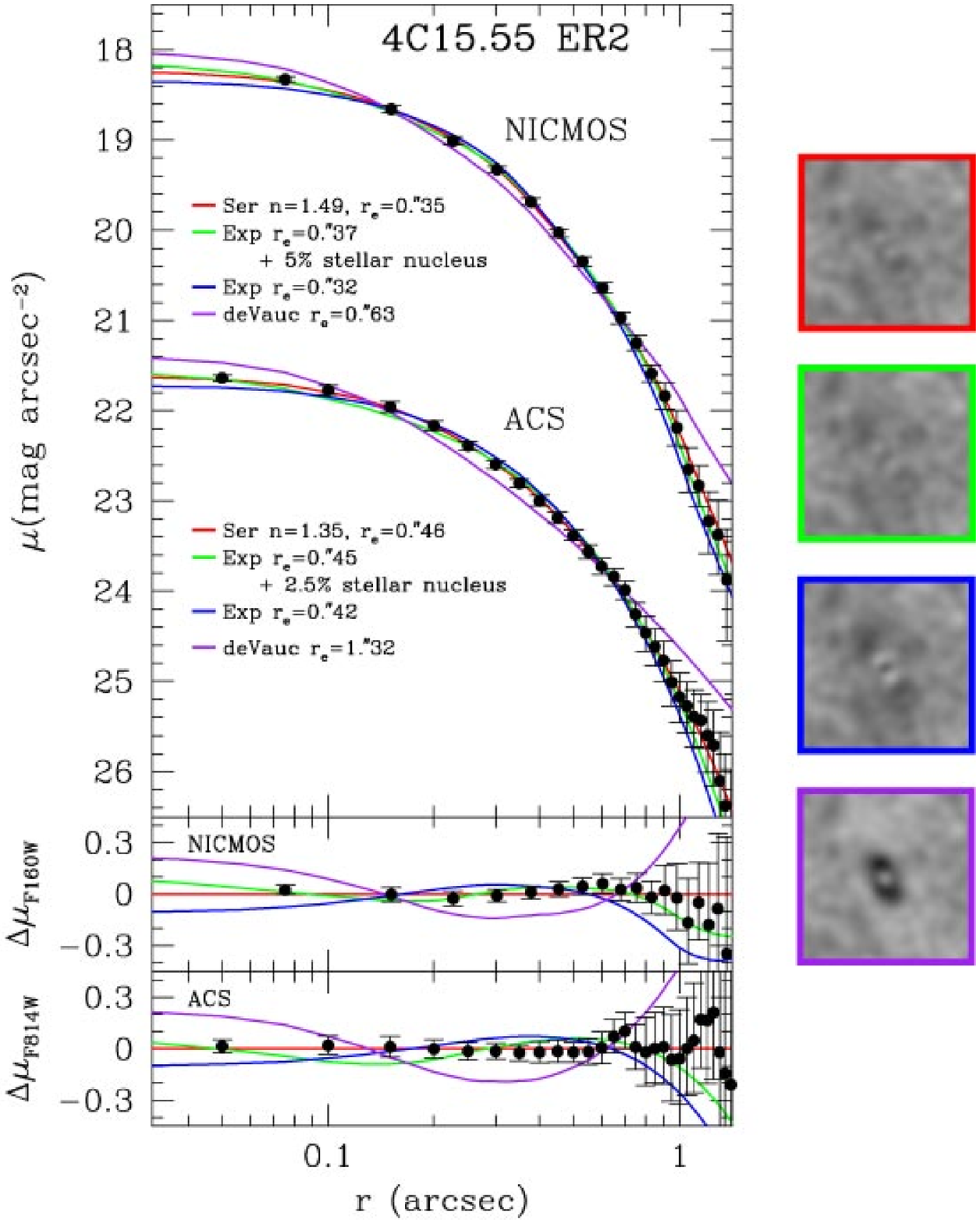}
\caption{Surface brightness profile for 4C 15.55 ER2.  
The best fit model is an exponential disk with a small point-like core that contributes between $2.5-5$\% of the total light. A S\'ersic profile with a low index ($n=1.49$) is also a good fit.  A pure r$^{1/4}$-law profile does not provide a good fit to either the NICMOS or ACS observations.  \label{4c15morph}}
\end{figure*}

From the HST imaging, 4C 15.55 ER2 appears from simple visual inspection to be either an elongated elliptical 
or an inclined disk galaxy.  Previous results from high S/N ground-based AO imaging concluded that the galaxy was a nearly pure disk galaxy \citep{sto06} 
with a small point-like core.  The NICMOS imaging corroborates this result.  The best {\sc galfit} model is an exponential profile with the addition of a central PSF that contributes 5\% of the total light (Fig.~\ref{4c15morph}).  A S\'ersic profile with an index, $n=1.5$ also fits the data well.  In either case, it is clear that this galaxy is dominated by a massive disk of old stars.  The ACS profile yields a similar result, with a somewhat smaller S\'ersic index ($n=1.35$) or an exponential disk with a slightly smaller core (contributing 2.5\% of the total light).  An r$^{1/4}$-law profile clearly provides a much worse fit to both the ACS and NICMOS imaging data.

\subsection{TXS 1812+412 ER2}

\begin{figure*}[htb]
\begin{center}
\epsscale{0.8}
\plotone{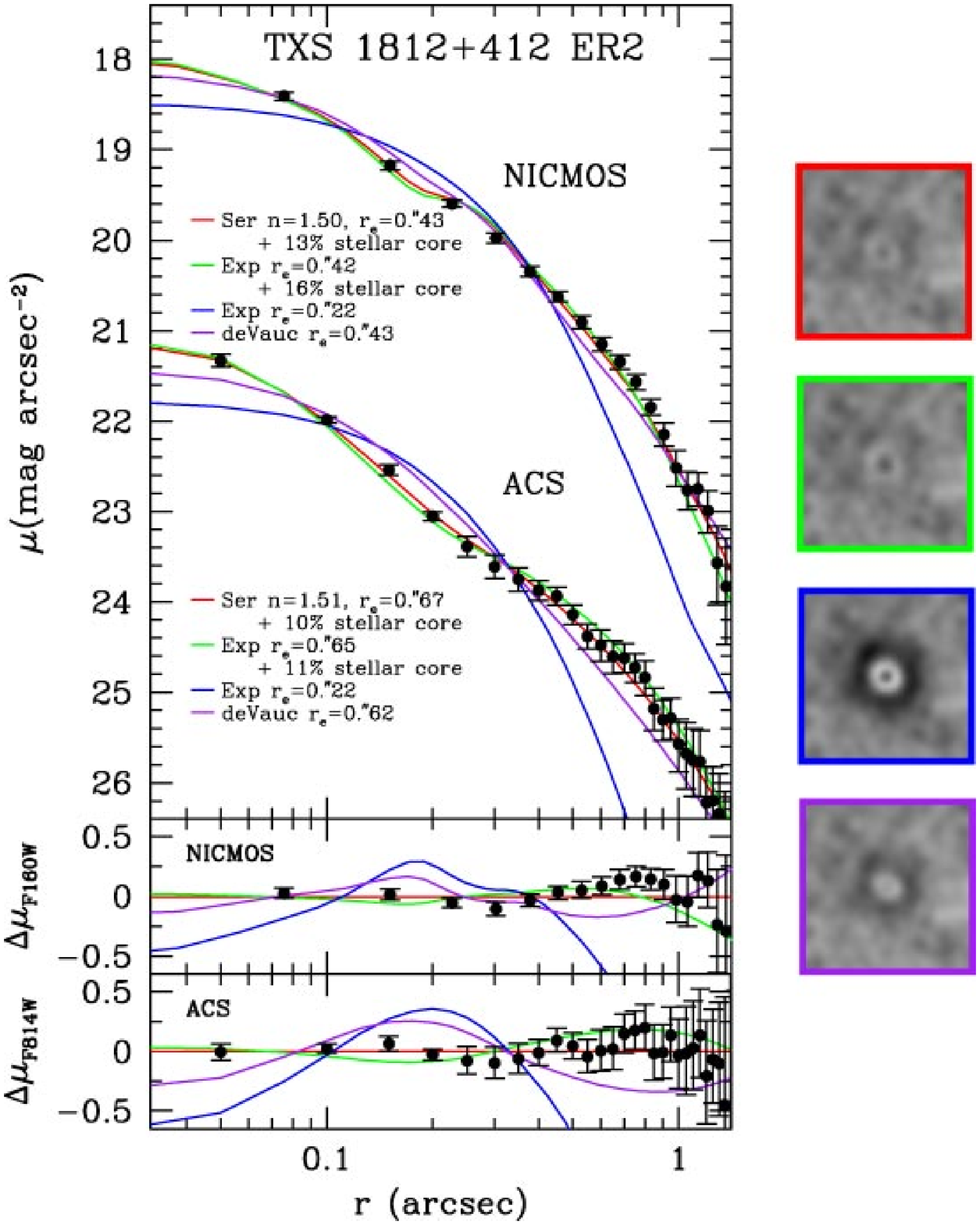}
\end{center}
\caption{Surface brightness profile for TXS 1812+412 ER2.  
The best fit for this galaxy is a two component model of either a S\'ersic profile with a low index ($n=1.50$) plus a point-like core that contributes 13\% of the light, or an exponential profile 
plus a small core that contributes $\sim$16\% of the light.  Disks with compact r$^{1/4}$-law bulges (not shown) provide nearly as good a fit as the stellar cores.
Pure exponential and pure de~Vaucouleurs models are ruled out.  \label{t1812morph}}
\end{figure*}

TXS 1812+412 ER2 appears to be fairly round with some extended diffuse emission around the core.  This fainter emission is nearly symmetrical in both the ACS and NICMOS images.  The best fit {\sc galfit} profile is a combination of something like a disk (an exponential profile or a S\'ersic profile with a low index) plus a small core (either point-like, or a small de~Vaucouleurs bulge).  Figure \ref{t1812morph} shows the fits from different single and multi-component models.  Pure exponential and de~Vaucouleurs profiles do not provide a good fit to the data as can be seen most clearly in the difference panels at the bottom of Figure \ref{t1812morph}.

\tabletypesize{\small}
\begin{deluxetable*}{rcccccc}
\tablecolumns{7}
\tablewidth{0pc}
\tablecaption{Properties of Old Galaxies}
\tablehead{
\colhead{Galaxy} & \colhead{$z$} & \colhead{best fit S\'ersic model} & 
\colhead{two-component model} & \colhead{$r_{eff}$} & \colhead{class} \\
\colhead{} & \colhead{} & \colhead{} & \colhead{} & \colhead{(kpc)} & \colhead{}
}
\startdata
TXS 0145+386 ER1 & 1.4533 & $n=4.3$ & \nodata & 6.65 & spheroid \\
ER2 & 1.459\tablenotemark{$\ast$} & $n=1.9$ \& $n=4.0$ & \nodata & 5.12 \& 0.77 & merger \\
J094258+4659.2 ERO R1 & 1.50\tablenotemark{$\ast$} & $n=3.6$ & expdisk + 32\% bulge & 7.34 & disk \\
B2 1018+34 ER2 & 1.4057 & $n=3.6$ & expdisk + 24\% bulge & 5.95 & disk \\
TXS 1211+334 ER1 & 1.598 & $n=3.9$ & \nodata & 2.22 & spheroid \\
4C 15.55 ER2 & 1.412 & $n=1.5$ & expdisk + 5\% PSF & 3.15 & disk \\
TXS 1812+412 ER2 & 1.290 & $n=1.5$ + 13\% PSF & expdisk + 16\% PSF & 3.63 & disk \\
\enddata
\tablenotetext{$\ast$}{The best-fit photometric redshift is given as no spectroscopic redshift is available for these two sources.}
\label{tab_prop}
\end{deluxetable*}

\section{Color Gradients}

Elliptical galaxies in the local universe exhibit color gradients, with the inner regions being redder than the outermost regions \citep{fra89, pel90}.  
This trend has been shown to be largely due to a metallicity gradient, rather than an age gradient
\citep{dav93, kob99, tam00, hin01}. The strong gravitational potential in the center of a galaxy contributes to the retention of high-metallicity gas, whereas in the outer regions of a galaxy, stellar winds from supernovae are efficient in expelling this metal-rich gas, leading to the observed color gradient.  
Such a gradient can be explained quite naturally with a monolithic collapse formation scenario, in which the bulk of the stars are formed in a single short burst and passively evolve thereafter.  Metallicity gradients in this scenario are likely in place shortly after formation, and remain roughly unchanged as the galaxy evolves.  
On the other hand, hierarchical formation scenarios predict that massive ellipticals form over longer periods of time, gradually being built up from smaller sub-units.  These models predict that elliptical galaxies consist of several distinct stellar populations, with mergers driving gas and dust into the central regions promoting mixing and fueling star formation which in turn dilutes color gradients (e.g., \citealt{whi80}), possibly even leading to inverted color gradients (e.g., \citealt{im01}). 
At high $z$, color gradients can be used to infer formation mechanisms since 
the galaxies are closer to their formation epoch and the youngest stars will be significantly bluer than the oldest stars.  

\begin{figure*}[htb]
\begin{center}
\epsscale{1.0}
\plotone{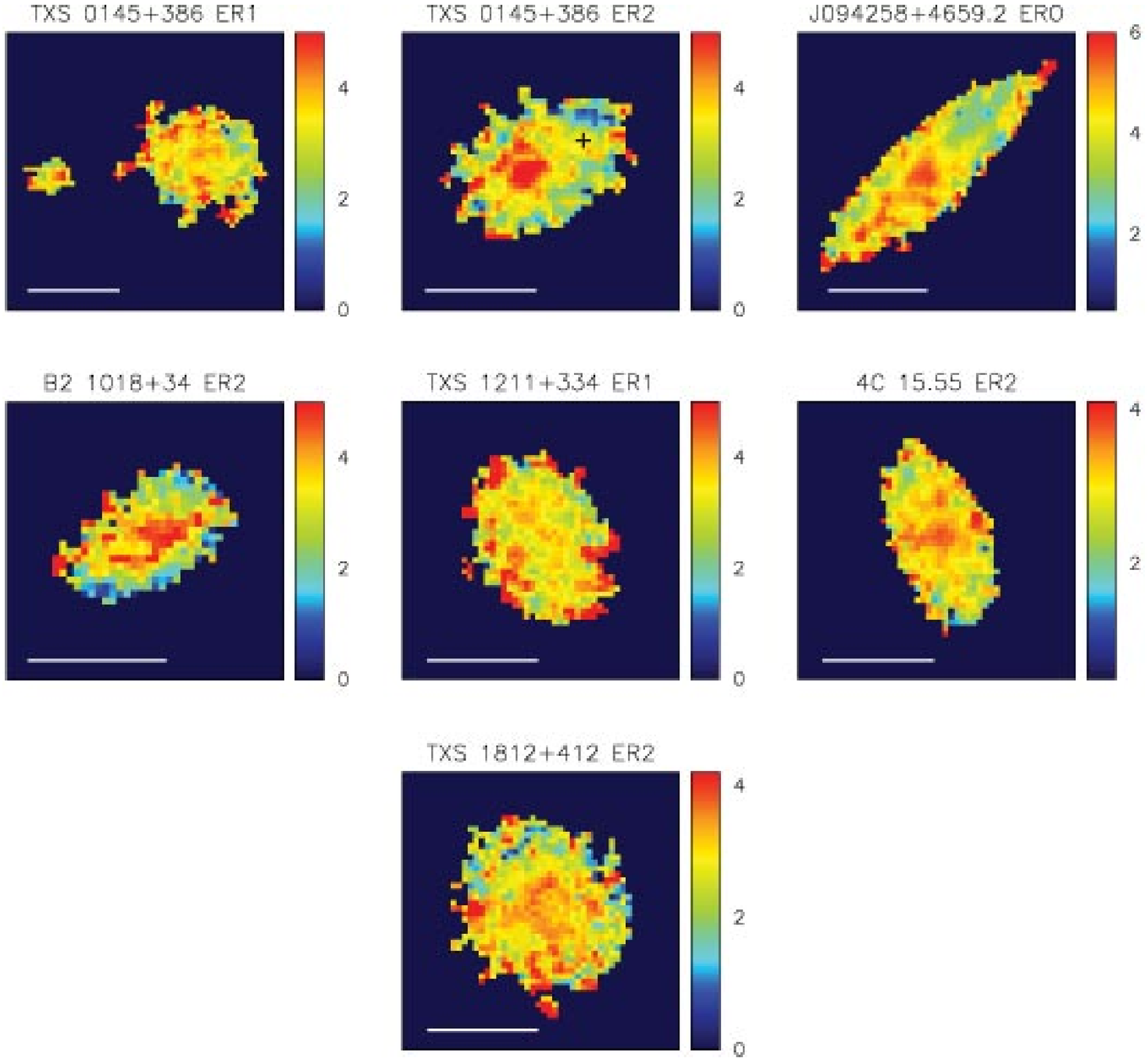}
\end{center}
\caption{Color maps for seven old galaxies from HST imaging.  TXS\,0145+386 ER2 consists of two galaxies; the centroid for the smaller galaxy in the merging pair is marked by a black cross.  
Scale for each image is denoted by a white bar which represents $1\arcsec$ in angular size.  Color wedges on the right hand side of each panel represent the F814W$-$F160W color scale for each corresponding galaxy in units of magnitude per pixel.
\label{color_fig}}
\end{figure*}

We have examined color gradients in our seven EROs to better understand what possible age or metallicity gradients are present.
In order to do this we first needed to resample the NICMOS and ACS images to put them on the same pixel scale, and then convolve them with the PSF from the other instrument.  
We used the same PSFs that were generated for the morphological modeling 
with the exception that the PSFs were resampled to the proper pixel scale before convolution.  This process eliminates any artificial gradient due to PSF differences between the F814W and F160W images.  We then rotated the images using header keywords that defined the telescope roll angle for each observation, and shifted them  using the galaxy centroids as a primary alignment.  When possible, other stars in the field were used to double check the rotation and offsets in order to ensure that differences in galaxy centroids between the blue and red components did 
not affect the alignment.  The background was subtracted from both images before
producing the final color maps.  We have made postage-stamp cut-out images of the galaxies using the SExtractor \citep{ber96} segmentation maps in order to mask out the surrounding noisy background, which fluctuates around zero.  These segmentation maps define the galaxy regions based on all contiguous pixels above a flux threshold of 2.5 $\sigma$ in the NICMOS imaging.  In terms of effective radii, r$_e$, the color maps generally include all galaxy flux within $\sim$2r$_e$, except for TXS\,0145+386 ER1, ER2, and B2\,1018+34 ER2 where this flux threshold retained all galaxy flux within 1.0 to 1.5 r$_e$.  These color maps are shown in Figure \ref{color_fig}.

The two galaxies that clearly follow r$^{1/4}$-law profiles, TXS 0145+386 ER1 and TXS 1211+334 ER1, have relatively flat colors, consistent with little or no color gradients.  These galaxies are uniformly red at least to $\sim1-2r_e$, where the S/N becomes too low to precisely determine any further color variation.  There is no evidence for any inverted
color gradient, such as that seen in roughly 50\% of the Hubble Deep Field galaxies \citep{men01}, 
which implies that these galaxies most likely have not suffered any recent 
star forming events.  This conclusion is corroborated by the individual ACS and NICMOS images, which show extremely relaxed systems with similar morphologies both short-ward and long-ward of the 4000{\AA}-break.  
However, the flatness of the color gradients in these two elliptical galaxies is quite different from what we observe in massive ellipticals in the local universe.  It is possible that some amount of dry merging has already occurred in these galaxies, diluting any initial metallicity gradients present from formation.  
It is interesting to note that a ``companion'' galaxy to the east of TXS 0145+386 ER1 has the same color as that of ER1.  This companion 
could be indicative of a possible future minor merger event.

The color map of TXS 0145+386 ER2, which consists of another merging pair, shows that one galaxy is clearly much redder than the other.  The observed color gradient in the main, red galaxy is most likely too steep to be reproduced by metallicity alone.  Furthermore, a blue tail appears to fan out to the northwest from the compact source, which could indicate a slightly younger disk component disrupted by the merger.  The combined colors of this merging pair yield an SED that is consistent with extremely little or no dust 
(Paper I).
If one of the two galaxies has a significantly younger population of stars, it would imply a much older age for the dominant stellar population in order to match the combined SED. 
Rest-frame UV spectroscopy, which was not obtained for this source, would be extremely helpful in determining whether any resulting star formation, however small, is occurring due to this merger, or whether it truly is a ``dry'' merger. 

B2 1018+34 ER2, which was determined in 
Paper I 
to be a younger, possibly star-forming galaxy has a very interesting color profile.  The central bulge is red, and a dust-lane is visible running in front of the bulge.  Bluer regions from the spiral arms visible in the ACS image are apparent on either side of the bulge.  Overall, it seems clear that this is a classic grand design spiral galaxy, with star formation occurring more recently in the spiral arms than in the central bulge region.  
However, the possible presence of [\ion{O}{3}] contaminating the $J$-band photometry (Paper I), may indicate that an AGN is partially responsible for the observed color gradient.

J094258+4659.2 ERO, discovered by \citet{iye00}, 
also shows an interesting color profile.  Results from the surface brightness profile fitting for this galaxy concluded that it was dominated by an exponential disk, with a bulge component that contributed $\sim$30\% of the light.  Here we can see clearly that one side of the galaxy is bluer than the other, probably indicating the presence of dust on the southeast side.  
The bulge is clearly red, and its stars are most likely slightly older than those in the disk.
If we take the color difference between the NW and SE sides of the disk and assume the difference arises solely from reddening effects, the extinction we infer based on the standard Galactic extinction curve ($R_V=3$) is consistent with the globally averaged extinction derived from the SED fitting \citep{iye03}.

TXS 1812+412 ER2 was determined to have a best-fit S\'ersic profile with $n=1.5$ plus either a small point-like or compact r$^{1/4}$-law core.
It appears to have a rather classic elliptical color gradient, however, with a red central region, becoming bluer in the outermost regions.  
This mild gradient could also be explained by a disk component that is only mildly younger than the central bulge.

Lastly, and perhaps most interesting, is the color profile of 4C 15.55 ER2.  This galaxy is quite clearly dominated by an old stellar population (1.8 Gyr; Paper I) and is also clearly dominated by an exponential profile (Fig.~\ref{4c15morph}).  The observed color gradient in this galaxy, however, is reminiscent of that found in local giant elliptical galaxies.  
This gradient is almost certainly due to a metallicity gradient, rather than an age gradient, given the fact that something very close to a single instantaneous burst of star formation appears to be required in order to fit the photometry.  There is little evidence from either the photometry or the spectroscopic data for any recent star formation occurring in this galaxy.
All signs indicate that this is an extremely relaxed, massive disk of old stars that must somehow evolve into something resembling a massive elliptical galaxy by the present day.

\section{Discussion}

\subsection{Massive Disks of Old Stars}
Morphological classes for EROs at high $z$ have largely been assumed to be correlated with their spectral type. 
That is, dusty starbursts are generally assumed to be disk-dominated, while galaxies with evolved stellar populations are assumed to be bulge-dominated.
We have found three cases, however, 
or roughly half of our sample, 
that exhibit exponential (or nearly exponential) profiles and are dominated by extremely old stars, plus a fourth galaxy with an exponential profile and somewhat younger stars.
\citet{yan04} find a similar result for their sample of EROs at $z\sim1$, with a little over half (7/12) of their absorption-line-only systems being disk-dominated (see also, \citealt{yan03}).  In fact, even among their galaxies that show small amounts of [\ion{O}{2}] emission, they note that the underlying stellar populations often appear to be quite old.
On the other hand, \citet{abr07} 
find that nearly all of their spectroscopically quiescent systems in the Gemini Deep Deep Survey are bulges, and on the basis of this conclude that morphology is a reasonable indicator of spectral activity.  Their approach to morphological classification, however, is quite different from that of \citet{yan03} 
or from our own approach.  They use a technique  based on an asymmetry index and a Gini index, which can be thought of as a generalized concentration index \citep{abr03}.  
The requirement for a galaxy to be classified as bulge-dominated under this system is a low asymmetry ($<0.15$) and a high Gini index ($> 0.50$).  This assumes that disk galaxies will be more asymmetric due to spiral arms and knots of star formation, and less centrally concentrated than their elliptical counterparts.  However, a disk of old stars can be quite symmetric, so there is reason to be cautious when using this classification system.
Furthermore, it is well known, as \citeauthor{abr07}\ point out, that the Asymmetry-Gini (AG) classification cannot distinguish between E and S0 galaxies.  

In fact, we have found that the 
inability of the AG method to distinguish between morphological classes appears to extend to 
galaxies of all types that are dominated by old stars.
 J094258+46592 ERO, 4C\,15.55 ER2, and TXS\,1812+412 ER2, which appear to be dominated by disks of old stars from our NICMOS surface brightness profile fitting, all have asymmetry indices near 0.05 and Gini indices near 0.50 (two of them having a Gini index slightly greater than 0.50 and one being slightly less than 0.50), as
determined by the {\sc morpheus}\footnote{\scriptsize{\sc morpheus} is available at
http://odysseus.astro.utoronto.ca/$\sim$abraham/Morpheus/} code
\citep{abr07}.  These values 
would place them into the \citeauthor{abr07}\ ``early type'' category, which is clearly a misclassification in terms of morphology.  We therefore have based our morphologies on radial surface brightness profiles, since this method does not assume anything about the distribution of starlight in the galaxies we are measuring, and appears to give the most robust results, especially when dealing with evolved stellar populations.

For the four galaxies in our sample that were determined to have strong exponential components to their radial light profiles, we examined their sizes and compare them to those of present-day disk galaxies.
J094258+4659 ERO and B2 1018+34 ER2 have effective radii of 7.3 and 6.0 kpc, respectively, which correspond to disk scale lengths of 4.4 and 3.6 kpc.  This is roughly consistent with average disks in the local universe \citep{van87} 
which have scale lengths of 5.5$\pm$2.7 kpc, and is comparable to the massive disks found by \citet{lab03} at $z>2$.  These galaxies could well be the passively evolved descendants of such a population.  The effective radius for 4C 15.55 ER2, on the other hand, is only 3.2 kpc, corresponding to a scale length of 1.9 kpc, and the effective radius of TXS\,1812+412 ER2 is 3.6 kpc, corresponding to a scale length of 2.1 kpc.  While the number statistics are low, taken together there is a general trend that these old disk galaxies at $z\sim$1.5 are 45\% smaller than present-day disks.  This is in line with recent work on the size evolution predicted by CDM dark matter halo growth \citep{som07}.  

While their size may be well explained under standard CDM predictions, the three oldest disk galaxies, J094258+46592, 4C\,15.55 ER2, and TXS\,1812+412 ER2, pose an interesting problem for hierarchical formation scenarios nonetheless.  4C\,15.55 ER2, in particular, is difficult to explain since it appears to be a nearly pure disk of 2$\times10^{11} M_{\sun}$ that completed star formation by $z>2.4$.  Any mergers that may have occurred between $1.4<z<2.4$ would have to have been minor mergers of gas poor systems in order to preserve both the disk structure and the age of the stellar population.  It therefore seems unlikely that mergers could be the dominant formation mechanism for such a galaxy.  Instead, something more similar to monolithic collapse, where star-formation proceeds rapidly and efficiently in a single burst spread more-or-less evenly across the disk, seems much more reasonable.

The selection of our sources near radio-loud quasars likely places them in more overdense regions that will eventually form present-day clusters.  As such, they are unlikely to be representative of all quiescent galaxies at $z\sim1.5$.  
However, the galaxies were chosen purely by photometric criteria, and there is no reason to believe that disks would be preferentially selected.  An equivalent selection in the local universe would result almost exclusively in elliptical and S0 (i.e., bulge-dominated) galaxies.  Therefore, while not necessarily representative of all quiescent galaxies at high redshift, it is clear that at least some massive galaxies are unlikely to have assembled the bulk of their stars through mergers.

\subsection{Size Evolution}
A number of recent studies have found evidence for highly compact, massive galaxies at $z>1.4$ (e.g., \citealt{tru06, zir07, tof07, van08}). However, the majority of our sources are not especially compact when compared with local galaxies of similar mass.  The typical effective radius of a $2\times10^{11} M_{\sun}$ galaxy in the SDSS is 6 kpc at rest-frame I-band \citep{she03}.  The galaxies presented here are typically $\sim$6 kpc at rest-frame R-band. Three galaxies (TXS\,1211+334 ER1, 4C\,15.55 ER2, and TXS\,1812+412 ER2) are more compact than this by a factor of $\sim$2.  Note, however, that we use the \citet{cha03} IMF which results in mass estimates that are a factor of 1.5 times smaller than masses calculated using the \citet{kro01} IMF as in the SDSS \citep{lon05}.  If we apply this correction factor to our mass estimates, then the typical size for galaxies of similar mass in the SDSS would be 7.0 kpc, and our galaxies would generally be 15 \% smaller than this, with three more than 50\% smaller than this. There is a slight tendency toward smaller effective radii at longer wavelengths (e.g., our rest-frame $R$-band vs. rest-frame $U$-band data), which could increase the discrepancy further if we had morphological information at rest-frame $I$-band to compare with the SDSS. However, in most cases size appears to be roughly independent of observed wavelength, as evidenced by the approximately flat color gradients seen in many of these galaxies. None of our galaxies are as compact as the $r < 1$kpc galaxies found at $z>2$ by \citet{tof07} and \citet{van08}.  Our three smallest galaxies are consistent with galaxies at $z\sim1.4$ in the \citet{tru06} sample, while the rest are significantly larger.  

It is interesting to note that the galaxies with the lowest dust content ($A_V<0.6$) from the photometric SED fitting (with the exception of TXS\,0145+386 ER2, which appears to be undergoing a dry merger), also appear to be the most compact, regardless of the disk or bulge-like nature of their morphology.  
This may be 
a further manifestation of the
relationship between star-formation and size at $z\sim2.5$ found by \citet{zir07} and \citet{tof07}.  
All of our galaxies are best fit by single, instantaneous bursts of star formation at high redshift rather than constant star formation models and would therefore all fall under the ``quiescent'' category of \citet{zir07} and \citet{tof07}.  However, the evidence for dust in galaxies with larger effective radii (and likewise, the lack of dust in the most compact sources) may point to a relationship between star-formation \emph{history} and size.  In this scenario galaxies that form rapidly and undergo intense, highly efficient star formation early-on end up as compact objects, either disks or bulges, while galaxies where star formation is slightly less efficient may leave behind more dust and may form more gradually, leading to extended morphologies.  
While the number statistics are low, this idea would appear to be supported by the fact that galaxies with similarly aged stellar populations at higher redshifts (e.g., \citealt{tof07,sto07,van08}) tend to have even more compact morphologies.  These $z>2$ galaxies must have formed over even shorter time periods than the galaxies in the present $z\sim1.5$ sample.

\subsection{Role of ``Dry Mergers"}
Regardless of how these galaxies formed originally, it is clear that given the age of the universe, over the remaining $\sim$12 billion years from the end of star formation until the present day, it is likely that they will undergo a number of subsequent mergers.  This is even more evident for the disks of old stars that we see at $z\sim1.5$, since we do not find any comparable galaxies in the local universe.  Some mechanism must be responsible for converting them into objects that more closely resemble either S0 or giant elliptical galaxies in the present-day.  Furthermore, the lack of luminous blue galaxies in the local universe requires that this mechanism must not destroy or dilute the old stellar populations present in these galaxies.

There has been much discussion recently about ``dry mergers", in which little or no cold gas is available to form new stars, as an important mechanism for preserving old stellar populations and generating the massive spheroids we see in the local universe \citep{bel04, van05, fab05}.  
\citet{bel04} find that the rest-frame B-band luminosity density of red galaxies evolves only mildly between $0<z<1.1$.  This requires at least a factor of two increase by $z=0$ in the stellar mass density present on the red sequence in order to compensate for dimming due to continued passive evolution of the stellar populations.  \citeauthor{bel04} attribute this evolution to both gas poor mergers and to truncation of star formation in blue star-forming galaxies.
\citet{bel06} also studied the dry merger fraction among galaxies in the GEMS survey and found that luminous early-type galaxies on the red sequence will undergo on average $\sim$1 major dry merger between $z<0.7$ and the present day.  
\citet{van05} finds a similar result, with $\sim$70\% of bulge-dominated galaxies experiencing a dry merger with a median mass ratio $1:4$ in the recent past ($z<0.2$).  He derives a current mass accretion rate of roughly 10\% Gyr$^{-1}$, which again implies 
about
a factor of 2 increase in stellar mass density in luminous red galaxies between $0<z<1$.

TXS 0145+386 ER2 may be 
an example of this dry merging process in action.  The colors of the galaxies in this merging pair  are consistent with that of old stellar populations rather than dusty starbursts (Paper I), and there is no evidence for any significant amount of recent star formation.  Recall that the compact, northwest galaxy in this pair was best fit by a very small effective radius, while the second, more extended galaxy to the southeast appeared to be best fit by something similar to an exponential disk, especially at rest-frame U-band.  In addition, the color map for this galaxy showed evidence for some possibly younger (although still relatively old) stars fanning out from the compact NW galaxy, which may imply that this spheroid is the central bulge of a disk galaxy being torn apart by the interaction.  If these scenarios are correct, this could imply a dry merger between two old disk galaxies, which would provide the much needed link between these oddities at $z=1.5$ and present-day massive ellipticals.

Another of our galaxies, TXS\,1211+334 ER1, was best fit by an r$^{1/4}$-law, but 
residuals along the axis of elongation suggest a more complex formation history.  It is possible that this galaxy suffered a merger in its recent past. However, given that no signatures of star formation are visible in the rest-frame UV spectrum of the source and that the SED favors a stellar population that is at least 1.4 Gyr old (Paper I) with a current star formation rate of $\leq0.2$ M$_{\sun}$ yr$^{-1}$ (Table \ref{tab-comp}), any merging event must have been gas poor.  
Taken together, TXS\,0145+386 ER2 and TXS\,1211+334 ER1 may represent different stages of a similar process.

\section{Summary and Conclusions}
We have studied a sample of seven evolved galaxies at $z\sim1.5$ with high resolution imaging from HST.  These are summarized below.
\begin{enumerate}
\item
\emph{TXS\,0145+386 ER1} is a relaxed elliptical galaxy at $z=1.4533$ with a dominant stellar population that is $\sim$1 Gyr old (Paper I).  There is no evidence for any color gradient across the galaxy, implying the stars are uniformly old, yet may have been mixed through dry merging events in the past.  A ``companion'' galaxy 2$^{\prime\prime}$ to the east  has the same color as ER1 and may indicate a current or future minor merging event.  There are no visible signs of interaction between these two galaxies, but these features may be too faint to observe at this redshift, especially if they are dominated by old stars.
\item
\emph{TXS\,0145+386 ER2} is a pair of merging galaxies at $z\simeq1.46$.  One galaxy is extremely compact, while the other is much more extended and redder in color.  
Their combined flux gives an SED of an extremely old stellar population (2.5 Gyr) with no dust (Paper I).  This result implies that no recent star formation is occurring and that they 
are undergoing a dry merger.  The ACS image of this pair shows some irregular structure in the extended galaxy which, if due to small amounts of star formation and/ or dust, implies even older ages for the majority of the stars in order to produce an average age of 2.5 Gyr.  There is some evidence from the color map, as well as a visible elongation in the ACS image, for bluer stars in a tail outward from the compact galaxy to the northwest; however the S/N is low in this outermost region.
\item
\emph{J094258+4659.2 ERO} is a disk-dominated galaxy at $z\simeq1.5$ with a bulge that contributes roughly 30\% of the total light.  The dominant stellar population is on average 1 Gyr old \citep{iye03}.  
A color gradient is present, with the central bulge being redder than the disk, and the disk is noticeably bluer on the northwest side and redder on the southeast side.  This is most likely due to slightly different stellar populations in the bulge and disk, as well as dust obscuration along the southeast side of the disk.
\item
\emph{B2\,1018+34 ER2} is a grand design spiral galaxy at $z=1.4057$ with a stellar population that is on average 0.64 Gyr old (Paper I), although it is likely that the majority of the stars are much older ($\sim1.9$ Gyr, Table \ref{tab-comp}) with a small contribution by mass from younger stars.  Spiral arms are most clearly visible in the ACS image, while the NICMOS image has a smoother profile.  A color map of the galaxy shows that a dust lane runs from the southeast to the northwest in front of the bulge.
\item
\emph{TXS\,1211+334 ER1} appears to be a classic elliptical galaxy at $z=1.598$ with an age of 1.4 Gyr (Paper I).  Residuals from subtracting an r$^{1/4}$-law profile from the NICMOS image indicate that the galaxy may have suffered a merging event in the past and may still be in the process of relaxing.  There is no evidence for any significant color gradient across the galaxy, which may be due to population mixing from prior merging events.
\item
\emph{4C\,15.55 ER2} is essentially a pure disk of old (1.8 Gyr, Paper I) stars at $z=1.412$.  A moderate color gradient is visible in this galaxy, with the central regions slightly redder than the outermost regions, which could indicate a metallicity gradient.  It is important to note that there does not appear to be a significant bulge component to this galaxy ($<5$\% of the total light), so this color gradient cannot be explained by a difference in stellar populations between a bulge and disk component.  
\item
\emph{TXS\,1812+412 ER2} appears to be another disk-dominated galaxy at $z=1.290$ with an age of 1.4 Gyr (Paper I).  The best-fit profile for this source requires a $\sim 10 - 15$\% contribution from a compact core.  There is a mild color gradient visible for this source, with the central regions being slightly redder than the outer regions.
\end{enumerate}

From the morphologies of these galaxies it is clear that age, or spectroscopic activity is not necessarily a surrogate for morphological class, and vice versa.  Three of these galaxies (J094258+4659.2, 4C15.55 ER2, and TXS\,1812+412 ER2) have strong exponential components to their radial light distributions, and both their broad-band SEDs and rest-frame UV continua point to evolved stellar populations (\citealt{iye03}; Paper I).  The existence of massive evolved disks at high redshift is a challenge for current semi-analytic models of galaxy formation, since it implies that they have not undergone 
a single major merging event since their formation more than 1 Gyr earlier.  Instead, they appear to have formed by a mechanism more similar to monolithic collapse.  Furthermore, their existence is problematic for new theories that rely on AGN feedback in galaxy formation in order to reproduce the observed large scale properties of galaxies, since powerful AGNs 
are likely triggered by major mergers and 
preferentially exist in luminous spheroids.  It would seem that another mechanism must be responsible 
either for triggering the AGN feedback or 
for terminating star formation in these disk galaxies at early times, possibly by the rapid formation and efficient processing of all available gas into stars in a single burst of activity.

Dry mergers may be the important link between these old galaxies and present-day massive ellipticals.  We have found at least one example of this process in action.
Dry mergers appear to be an important evolutionary phase in most luminous red galaxies 
at $z<1$ \citep{bel04, van05, bel06}.  However, massive disks of old stars that have not
suffered a single major merger are present at early times, implying that multiple formation mechanisms may be required to explain how the earliest massive galaxies formed.
It will be important to continue morphological studies of old galaxies at redshifts $z>1.5$ (e.g., \citealt{sto07}) in order to determine how common massive disks of old stars are in the early universe.

\acknowledgments

We would like to thank the anonymous referee for useful suggestions that helped improve the paper.  This research was partially supported by NSF grant AST 03-07335 and by grant GO-10418 from the Space Telescope Science Institute, which is operated by
the Association of Universities for Research in Astronomy, Inc., under NASA contract
NAS 5-56555.

%\clearpage

\end{document}